# RIP Analysis of Modulated Sampling Schemes for Recovering Spectrally Sparse Signals

Ray Maleh, Gerald L. Fudge
L-3 Communications Mission Integration

*Abstract*— **In this work, we analyze modulated sampling schemes, such as the Nyquist Folding Receiver, which are highly efficient, readily implementable, non-uniform sampling schemes that allows for the blind estimation of a narrow-band signal's spectral content and location in a wide-band environment. This non-uniform sampling, achieved by narrow-band modulation of the RF instantaneous sample rate, results in a frequency domain point spread function that is between the extremes obtained by uniform sampling and totally random sampling. As a result, while still preserving structured aliasing, the modulated sampling scheme is also useful in a compressive sensing (CS) setting. We estimate restricted isometry property (RIP) constants for CS matrices induced by such modulated sampling schemes and use those estimates to determine the amount of sparsity needed for signal recovery. This is followed by a demonstration and analysis of Orthogonal Matching Pursuit's ability to reconstruct signals from noisy non-uniform samples.**

## I. INTRODUCTION

Thanks to the work of Nyquist [1], it is a well known fact that a band-limited signal can be recovered by sampling uniformly at a rate twice its bandwidth. In the case of bandpass sampling, a relatively slow speed analog-to-digital converter (ADC) samples an RF band and creates an "alias" of the original signal's spectrum that appears in the frequency region bounded by one-half of the sampling rate [2]. While the signal can be reconstructed, it is impossible to determine the signal's original carrier frequency without *a priori* information such as knowledge of an anti-aliasing filter. If the anti-aliasing filter is removed so that a wideband RF region is sampled at Nyquist for the *n*th RF band [$n f_S/2$, $(n+1) f_S/2$] (or Nyquist zone) but sub-Nyquist for the total bandwidth consisting of multiple Nyquist zones, we get aliasing from multiple Nyquist zones simultaneously, an outcome typically to be avoided. But, if we can sample in such a way that the signal spectral "aliases" are not identical and can be distinguished from one another, then we can exploit this encoded information to recover the signal's original frequency location. A graphical depiction of this concept is illustrated in Fig. 1 where the baseband Nyquist zone (DC to $f_S/2$) is highlighted in tan on the left portion of each spectral case. The top panel shows the original RF signal in the $z = 3$ Nyquist zone prior to RF sampling. The middle panel shows identical aliasing, such as the case with conventional uniform sampling, while the bottom panel shows aliased images that can be distinguished from each other.

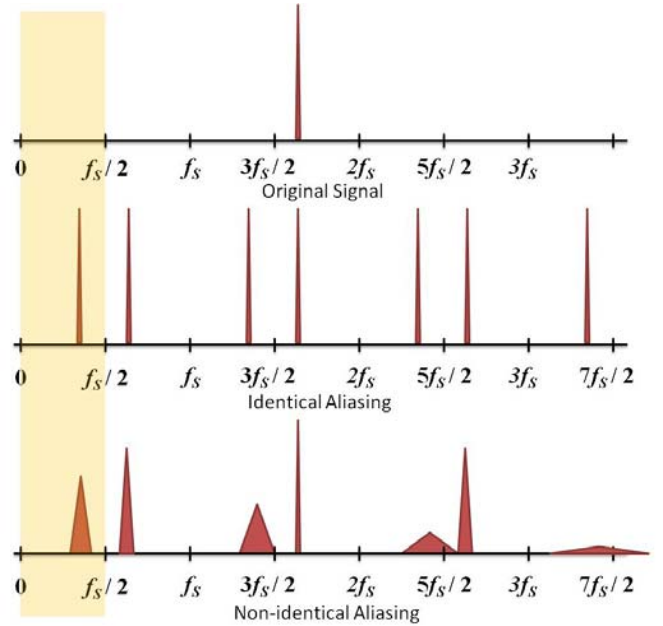

Fig. 1. Identical and Non-Identical Aliasing

There are currently several sampling architectures that induce non-identical aliasing: Multi-coset Sampling [3] utilizes multiple clocks that are non-uniformly interleaved. Along similar lines, the Modulated Wide-band Converter (MWC) [4],[5] incorporates multiple clocks that use a rapidly fluctuating periodic signal (as opposed to a Dirac pulse stream) to induce aliases that differ in magnitude. Through a dimensionality reduction, the multi-coset sampling and the MWC can recover several CW tones from a far fewer number of sample clocks. While highly effective at producing a solution, both multi-clock architectures suffer when hardware

This work was supported in part by the Air Force Research Laboratory through the DARPA Analog-to-Information Program under AFRL contract FA8650-08-C-7852. Approved for Public Release, Distribution Unlimited. The views expressed are those of the author and do not reflect the official policy or position of the Department of Defense or the U.S. Government.

The authors are with L-3 Communications Integrated Systems, Mission Integration Division, Greenville, TX 75402, USA (e-mail: Ray.Maleh@L-3com.com, Gerald.L.Fudge@L-3com.com)

complexity and power consumption are taken into account. Ideally, from a resource conservation point of view, we would like to limit our system to only one clock. Other systems such as that in [6] modulate a signal by a random P-N chip sequence that effectively modulates a sparse-in-frequency signal by a highly randomized transfer function. In this case, aliasing is extremely non-uniform. Every CW tone contained within the signal has a unique signature within some prescribed base-band region, thus making it possible to recover the original signal using compressive sensing techniques. While theoretically promising, this technique requires the design of a circuit that toggles between two voltages according to a pseudo-random sequence at the Nyquist rate. This is a highly non-trivial task. In addition, reconstruction is computationally expensive.

Another architecture that induces non-identical aliasing is the Nyquist Folding Receiver [7], or NYFR, which is the focus of this effort. The NYFR introduces phase modulation onto an RF sample clock local oscillator in order to produce a non-uniform sample pulse train. The resulting non-uniformly spaced sample points will induce non-uniform aliasing, from which a narrow-band signal's original RF can be derived by examining the spectral content of the Nyquist base-band, i.e. the interval $[0, f_S/2)$ where $f_S$ is the average sample rate. This is illustrated in Fig. 1 and Fig. 2.

There are a number of approaches to recover information from signals that are sampled via this architecture including spectrogram inspection or analysis, chirp slope estimation, matched filter hypothesis testing, quadrature mirror filtering, and de-modulation of the induced modulation [7],[8],[9]. In this paper, we will examine the information recovery problem from a compressive sensing [10] point of view.

In section II, we describe our modulated sampling architecture in greater detail and set up the related compressive sensing framework necessary to analyze our method. Then, in Section III, we use a simple example involving uniform sampling to outline the tradeoff between uniform sampling and highly randomized sampling. It turns out that the modulated sampling scheme is a highly effective compromise between these two extremes. This discussion will motivate the statistical analysis of the restricted isometry property in Section IV. This section will be useful for analyzing sampling schemes where aliasing is minimally suppressed. In cases where the induced clock modulation significantly attenuates a signal's aliased spectrum while still preserving its structure, we may obtain an exact RIP which will be derived in Section V. In Section VI, we present empirical evidence justifying Orthogonal Matching Pursuit's ability to recover sparse signals in a noisy environment. We compare OMP failure probabilities with optimal probabilities derived from the Cramér-Rao bound associated with chirp slope estimation. This is followed by an empirical estimation of RIP constants in Section VII. This work will show both theoretically and empirically that signals sampled according to a modulated sampling scheme can effectively be recovered via compressive sensing methods.

## II. BACKGROUND INFORMATION

Let $x \in X$ be a signal from the class of real-valued wide-band differentiable signals with compact support. We can obtain samples of the signal $x$ using the operator $S: X \to \mathbb{R}^K$ that returns the values of $x$ at time instants $\{t_k\}_{k=1,\cdots,K}$. For any signal $x$, we can define the coset $\langle x \rangle \in X/\ker(S)$ to be the set of all signals that have the same sample values as $x$. Every coset contains a unique base-band signal which is the element of the coset whose first spectral moment is closest to zero (i.e. DC). More formally, for any signal $x$, we define:

$$\mu(x) = \frac{2}{\|\mathcal{F}\{x\}(\omega)\|} \int_{-\infty}^{\infty} \omega |\mathcal{F}\{x\}(\omega)|^2 d\omega$$

Next define the map $b: X/\ker(S) \to X$ as follows:

$$b(\langle x \rangle) = \operatorname*{argmin}_{u \in \langle x \rangle} \mu(u)$$

Any ties can be broken using some choice function. The output of $b$ is the base-band representative of the input coset. We can also define the map $h = b \circ \pi : X \to X$ to be the function that returns the base-band representation of a given signal. Here, $\pi$ is simply the canonical projection of $X$ onto $X/\ker(S)$. For the case of a sampling scheme $S$ that is only mildly non-uniform, we can pictorially depict the above abstract frame as shown in Fig. 2. Observe that for a given narrow-band signal $x$ and a mean sample rate $f_S$, all we can "see" at the output of the interpolation filter is $h(x)$, that is, we only know $x$ up to its coset. In order to recover $x$ itself, we must examine additional structure inherent within each coset, which is precisely what our modulated sampling scheme does.

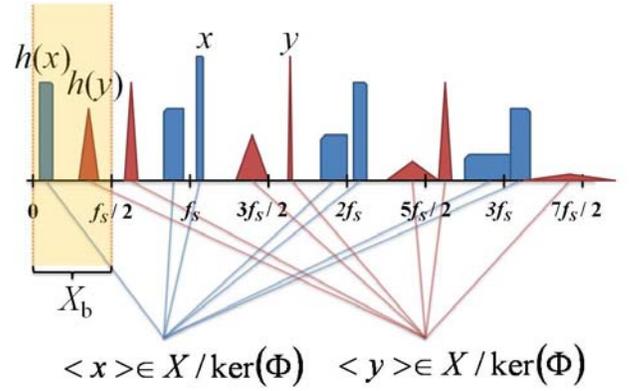

Fig. 2. Spectral schematic demonstrating our abstract framework.

The basic mathematical setup of the modulated sampling scheme is as follows. We sample $x$ with a pulse train as shown in Fig. 3. The sample times $t_k$ are the positive slope zero-crossings of an RF sample clock that may be modeled as a narrowband phase modulated signal centered at a carrier frequency $f_{S1}$ representing the average sample rate. The sampled signal is then passed through a low-pass filter with bandwidth typically equal to roughly half of the mean sampling frequency: this filter both serves to interpolate the sampled pulse stream and acts as an anti-aliasing filter for the conventional ADC, which runs at rate $f_{S2}$. We then perform information recovery after uniformly sampling with the ADC.

Note that the uniform ADC sample rate is de-coupled from the non-uniform RF sample rate by the interpolation filter.

We assume that the phase-modulated RF sample clock takes the form:

$$s(t) = \sin(\omega_{S1} t + \theta(t)) \quad (1)$$

It can be shown (see [7]) that if the input is a narrow-band signal of the form $x(t) = \cos(\omega_C t + \psi(t))$, then the output $y = h(x)$ of the system, before the ADC, will be:

$$y(t) = \cos(|\omega_C - k_H \omega_{S1}|t + \beta\psi(t) - M\theta(t))$$

where $M = \beta k_H$, $\beta = \text{sgn}(\omega_C - k_H \omega_{S1})$, and $k_H = \text{round}(\omega_C/\omega_{S1})$. We note that the value of $M$ depends on the Nyquist zone containing $x(t)$. For Nyquist zones $n = 0, 1, 2, 3, 4, 5, 6, \ldots$, it can easily be shown that $M = 0, -1, 1, -2, 2, -3, 3$, etc. Thus, the output of the receiver is the original folded signal plus an integer scale factor $M$ on the induced modulation with $M$ related in a one-to-one manner with Nyquist zone. This makes our system a highly suitable architecture for blind wide-spectrum estimation.

A two-signal example is shown in the bottom half of Fig. 3, with one signal in the $n = 2$ Nyquist zone and another signal in the $n = 3$ Nyquist zone. Items A and B correspond to the original spectra of two narrow-band signals. Item C illustrates the result of sampling using a modulated clock. Observe that as the original spectra move through the various Nyquist zones, their aliased copies vary in bandwidth. After applying an interpolation filter (item C), the bandwidths of the remaining spectra contain information that can be utilized to reveal the Nyquist zones of the original signals.

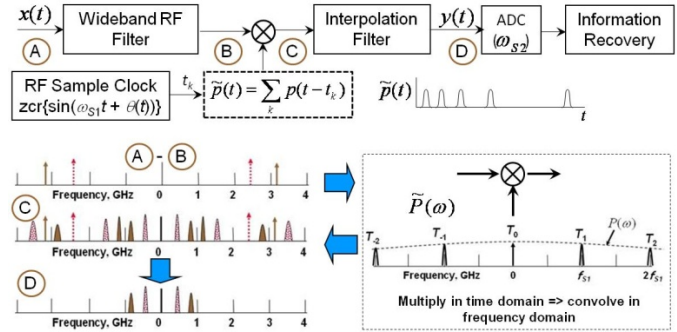

Fig. 3. Basic Modulated Sampling Receiver Architecture

We now consider the relationship between modulated sampling and the theory of compressive sensing. We first discuss the connection thoroughly in the discrete setting and then extend our results to signals of a continuous variable later. Let $\hat{x} \in \mathbb{C}^N$ be an $s$-sparse vector (consisting of $s$ non-zero entries) that represents the wide-band spectrum of the signal $x$. In such a case, we say that the signal $x$ is $s$-sparse in frequency. Let $y \in \mathbb{C}^K$ denote $K$ samples of the signal that are collected at the output of the ADC. We can relate $y$ to $\hat{x}$ via the underdetermined linear system

$$y = \Phi \hat{x} \quad (2)$$

The sensing or observation matrix $\Phi$ is a $K$ by $N$ sub-matrix of the $N$ by $N$ inverse DFT matrix where each row corresponds to a sample time determined by the values of $t_k$. This algebraic model is shown in Fig. 4 as a visualization of (13)-(15) in [11].

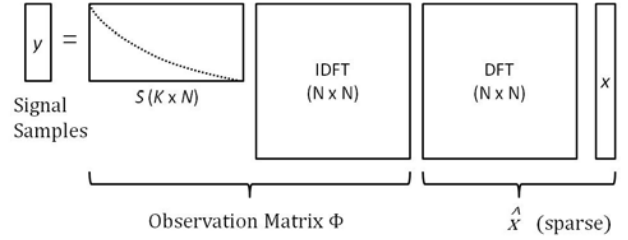

Fig. 4. Modulated Sampling Compressive Sensing Model

Here, the vector $x$ can be thought of as samples of the wide-band signal taken at or above the Nyquist rate relative to the full bandwidth RF input. Our sampling operator $S$ is a binary 0/1 matrix where each row $k$ has exactly one 1 in a column that corresponds to the $k$th sample time relative to the underlying sample grid. Of course, the actual sample times may not fall nicely on a grid, thus, the sample times may have to be approximated. For now, we will assume that the sample times do fall exactly on a grid. Later we will consider the limiting case where the sample grid is allowed to become arbitrarily fine in resolution in order to accommodate signals of a continuous variable.

Because $\dim(\ker(S)) \geq N - K$, Equation (2) is, in general, ill-posed and has multiple solutions; however, if $\hat{x}$ is sufficiently sparse, then under certain constraints on $S$ (and therefore $\Phi$), it is the unique sparse solution to this system. Intuitively, such constraints should ensure that the sampling matrix $S$ only admits cosets in $X/\ker(S)$ that have precisely one signal with a sufficiently sparse spectral representation. One way to enforce such a constraint is by means of the restricted isometry property (RIP):

**Definition 1:** A (general) compressive sensing matrix $\Phi$ has the restricted isometry property of order $s$ if there is a constant $0 \leq \delta_s < 1$ such that for *every* $s$-sparse signal $\hat{x}$:

$$(1 - \delta_s)\|\hat{x}\|_2^2 \leq \|\Phi\hat{x}\|_2^2 \leq (1 + \delta_s)\|\hat{x}\|_2^2$$

If we let $\Phi_\Lambda$ denote the submatrix of $\Phi$ obtained by selecting columns of $\Phi$ indexed by $\Lambda$, then we can see that an equivalent formulation of the RIP would be to say that all eigenvalues of matrices of the form $\Phi_\Lambda^* \Phi_\Lambda$, where $|\Lambda| \leq s$, sit within the interval $(1 - \delta_s, 1 + \delta_s)$. Thus, the RIP is a measure of how unitary the matrix $\Phi$ behaves with respect to sparse signals. Now if $\delta_{2s} < 1$, it follows that any coset in $X/\ker(S)$ can have no more than one signal with spectral support less than or equal to $s$. To see this, suppose that $x$ and $x'$ are two signals that are $s$-sparse in frequency and lie in the same coset. Then, $e = x - x' \in \ker(S)$ is $2s$-sparse in frequency. Thus, by the RIP, we have $0 = \|\Phi\hat{e}\|_2^2 \geq (1 - \delta_{2s})\|\hat{e}\|_2^2 \geq 0$, which implies that $\hat{e} = 0$. This implies that $x = x'$.

Compressive sensing problems of the form (2) can be solved by convex optimization methods as suggested in [12],[13], etc. Such approaches typically take the form:

$$\tilde{x} = \underset{z}{\operatorname{argmin}} \|z\|_1 \quad \text{subject to} \quad y = \Phi z \quad (3)$$

In the case of noisy samples or jitter (or if the sample times do not fall exactly on an underlying grid), the equality constraint in (3) can be replaced with an inequality constraint of the form $\|y - \Phi z\|_2 < \varepsilon$ for some noise tolerance parameter $\varepsilon$.

In [12], Candes proves the following fact regarding convex optimization's ability to recover a signal from noisy measurements:

**Theorem 1 (Candes):** Let $\Phi$ be a sensing matrix such that $\delta_{2s} < \sqrt{2} - 1$. Then the convex program (3) will recover any s-sparse signal $\hat{x}$ from its measurements.

Of course, convex optimization routines tend to be slow and, are therefore ill-suited for applications that involve fast processing of extremely large sampled data sets. Fortunately, there are several alternative fast iterative algorithms that solve the same compressive sensing problem (2). These include SpaRSA [14], Orthogonal Matching Pursuit (OMP) [15], CoSAMP [16], and Iterative Thresholding [17],[18],[19]. All of these algorithms have, at least empirically, been shown to perform as well as convex optimization.

This paper focuses on Orthogonal Matching Pursuit because of its efficiency and ease of implementation. In addition, OMP does have performance guarantees based on the restricted isometry property. The following theorem is proved in [20].

**Theorem 2 (Maleh):** Let $\Phi$ be a sensing matrix such that

$$\delta_{s+1} < \frac{1}{1+\sqrt{s}} \quad (4)$$

Then, OMP will recover any *s*-sparse signal $\hat{x}$ from its measurements.

While this recovery guarantee for OMP is weaker than convex optimization recovery guarantee in Theorem 1, it will be useful later when analyzing OMP's ability to identify two tones in a high dynamic range environment.

### III. UNIFORM SAMPLING AND RANDOM SAMPLING

We will now discuss uniform sampling in the context of aliasing and the RIP. This discussion is motivated by the fact that in shallow modulated sampling schemes, the RF sample clock phase modulation function $\theta(t)$ is narrowband and thus the sample times are almost uniformly spaced, so that the receiver output consists of a set of nearly-identical aliased images folded into the baseband Nyquist zone.

Unfortunately, the definition of the restricted isometry property is not directly compatible with nearly uniform sampling schemes. Such schemes purposely induce aliasing in order to recover high frequency data at a convenient intermediate frequency within a prescribed Nyquist zone. In the context of compressive sensing, a uniform sampler corresponds to an extremely poor Fourier sensing matrix. In CS, aliasing of any form is highly undesired. As a result, CS typically relies on highly random non-uniform sampling schemes which minimally exhibit any sort of periodicity that could translate into aliasing. To see this, consider Fig 5.

The image in the top left corner is the spectrum of a complex exponential signal of length 1024. The only non-zero frequency is at position 192. The plot in the top right corner shows the resulting spectrum when the original signal is sampled uniformly with sampling interval 8. Notice that the spike is periodically repeated every 128 units. Now if we sample the original signal with slight non-uniformity, we obtain the plot in the lower left hand corner. This sampling pattern was created by mildly perturbing the previous uniform sample points by random Gaussian "jitter" noise. Observe that the resulting spectrum has strong aliasing spaced out by 128 units in addition to "random" aliasing induced by the non-uniformity. Now finally, the plot in the lower right hand corner demonstrates the resulting spectrum when the signal is sampled completely randomly according to compressive sensing guidelines for guaranteed reconstruction. All aliasing is suppressed and the original spike is clearly present without ambiguity.

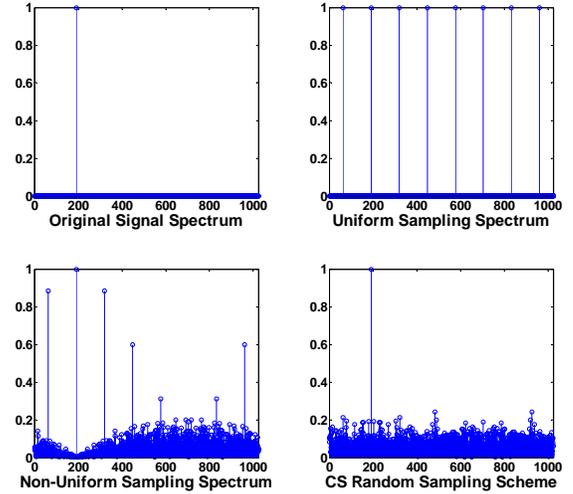

Fig 5: Examples of Impulse Responses of Various Sampling Schemes

It is clear that there is a tradeoff between the preservation of folded spectra and the suppression of aliasing. A uniform sampler will have poor RIP qualities. Conversely, a completely random sampling scheme will typically suppress all aliasing and have excellent RIP qualities. With modulated sampling, we find ourselves somewhere in the middle. In order for a modulated sampling scheme to theoretically perform well in a CS setting, the RF clock modulation must induce a sufficient amount of non-uniformity so that the RIP is satisfied; however, the modulation must not be so great that the aliased spectra fall below the noise floor.

Interestingly enough, despite the poor restricted isometry numbers that result from using Fourier dictionaries induced by uniform sampling grids, compressive sensing algorithms can still succeed with high probability. The restricted isometry number provides us with a worst case scenario. In other words, performing RIP analysis is equivalent to finding the worst possible *s*-sparse in frequency signal *x* such that

$$\|\Phi\hat{x}\|_2^2 = (1 - \delta_s)\|\hat{x}\|_2^2$$
$$\text{OR}$$
$$\|\Phi\hat{x}\|_2^2 = (1 + \delta_s)\|\hat{x}\|_2^2$$

Fortunately, such signals *x* occur with extremely low probability. To analyze this situation more closely, for any signal *x*, define the *spectral norm-deviation* of *x* as:

$$\delta_s(x) = \left| \frac{\left\|\Phi^*_{\Lambda(\hat{x})}\Phi\hat{x}\right\|_2}{\|x\|_2} - 1 \right| \quad (5)$$

where $\Lambda(\hat{x}) = \text{supp}(\hat{x})$. Observe that the restricted isometry number can be defined as a function of the spectral norm-deviation using the following equation:

$$\delta_s = \max_{\substack{x \text{ such that} \\ \hat{x} \text{ } s\text{-sparse}}} \delta_s(x) \quad (6)$$

Now consider the following simple example: Given a discrete uniform sampling setup with an atomic sampling frequency of 100 GHz, an RF sample clock running at 2 GHz, an observation window of 10 μs, and a signal *x* with two tones of amplitude $A_1$ and $A_2$ at frequencies $f_1$ and $f_2$ respectively, it is possible to show that:

$$\delta_2(x) = \begin{cases} \left|\sqrt{2 + \frac{4A_1 A_2}{A_1^2 + A_2^2}} - 1\right| & \text{if } f_1 \equiv f_2 \text{ (mod 20000)} \\ 0 & \text{otherwise} \end{cases}$$

We can obtain a lower bound on $\delta_2$ by setting $A_1$ equal to $A_2$ and observing that

$$\delta_2 \geq \left|\sqrt{2 + \frac{4A_1^2}{A_1^2 + A_1^2}} - 1\right| = 1$$

This uniform sampling scheme suffers from an extremely poor RIP constant that will not guarantee the success of any compressive sensing algorithm in discriminating between the two tones. However, one should ask the question: how often do signals with poor spectral norm-deviations occur? In the above example, $\delta_2(x)$ will be identically zero with probability at least 0.99995. Thus, CS methods should be able to recover practically every signal that is *digitally* 2-sparse in frequency. This corresponds to a recoverable bandwidth of approximately 200 kHz.

This simple example illustrates that while a sampling scheme may not allow for the recovery of a few "worst case" signals, in general, most sufficiently sparse signals can be reconstructed. This observation motivates the next section where we will apply a slightly weaker statistical analysis to derive sufficient conditions for the recovery of sparse signals using general sampling schemes.

## IV. STATISTICAL ANALYSIS OF THE RIP

To analyze any general sampling schemes, we can appeal to the Statistical Restricted Isometry Property (STRIP) as presented in [21]. The main result of [21] is as follows:

Let $\Phi$ be a fixed sensing matrix with *K* rows and *N* columns. Further assume that

1. The columns of $\Phi$ form an algebraic group under pointwise multiplication.
2. The rows of $\Phi$ are orthogonal and have zero sum.

Now let $\hat{x}$ be an *s*-sparse signal with non-zero entries whose locations are uniformly distributed. Let $\delta_s$ be a constant such that $(s - 1)/(N - 1) < \delta_s < 1$. Then we have

$$(1 - \delta_s)\|\hat{x}\|_2^2 \leq \|\Phi\hat{x}\|_2^2 \leq (1 + \delta_s)\|\hat{x}\|_2^2$$

with probability

$$1 - \frac{\frac{2s}{K} + \frac{2s + 7}{N - 3}}{\left(\delta_s - \frac{s - 1}{N - 1}\right)^2}. \quad (7)$$

It is not difficult to show that *any* Fourier sampling scheme, whether uniform or not, satisfies conditions 1 and 2 above. Thus, we can apply this result to the analysis of the modulated sampling scheme as follows: Assume $N = 1{,}000{,}000$ and $K = 20{,}000$ as before. Using these parameters, we may use (7) in order to calculate the maximum sparsity guaranteed to be recoverable by convex optimization for various failure probability tolerances as tabulated below in Table I.

**Table I: Iterative Greedy Unfolding**

| Failure Probability | Recoverable Sparsity |
|---|---|
| 0.1 | 84 |
| 0.05 | 42 |
| 0.01 | 8 |
| 0.005 | 4 |

The results generated by the STRIP analysis give us a theoretical framework with which we can analyze general signals without making additional underlying assumptions regarding dynamic range and/or coefficient decay rates.

For uniform and slightly perturbed/jittered sampling schemes this is the best that we can do. In such situations, the majority of spectral data in a signal gets shifted predictably according to Nyquist's theorem. Therefore, there is always some probability of a significant collision. On the other hand, for highly randomized sampling schemes, spectral energy spreads out randomly and the exact RIP results of [22] apply. For deterministic constructions of such highly random matrices, one can refer to [11] or [23]. Our interest lies at the median of these two extremes. If we use an optimal amount of sample clock modulation, we can significantly attenuate the aliasing while still preserving its structure. In a high SNR environment, we can exploit this remaining structure to obtain the exact RIP analysis of the next section.

## V. RIP ANALYSIS OF MODULATED SAMPLING SCHEMES

When modulation is introduced to the RF sample clock, the original spectrum of the signal remains unchanged; however, aliased spectra are blurred in frequency space. In other words, these ghost spectra will have increased bandwidths and decreased amplitudes. The magnitude of this effect is related to the number of Nyquist zones between an aliased spectrum and the original spectrum. As an example of this, consider Fig. 6 which is a plot of the folder spectra of four tones (500 MHz, 2.5 GHz, 4.5 GHz, and 6.5 GHz). Here, the RF sample clock is running at 2 GHz with a +/-100 MHz deviation induced by a modulated saw-tooth chirp.

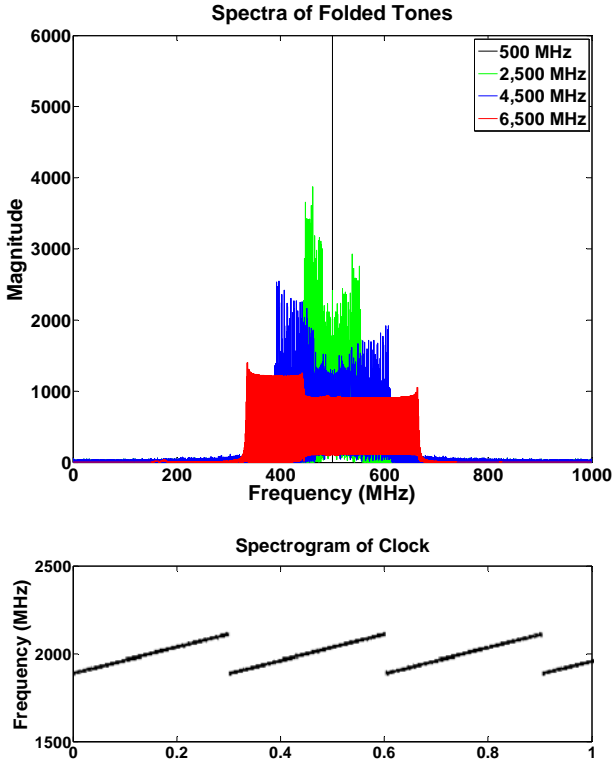

Fig. 6. Example 4-tone folded output spectrum

The tone at 500 MHz did not fold and thus has nearly zero bandwidth. The tone at 1.5 GHz folded from Nyquist zone 2 to 0. Thus, its folded spectrum has bandwidth $1 \times 100 = 100$ MHz. Observe that as the bandwidth increases, the magnitude of the spectra decrease. In fact, it seems to be the case that despite the bandwidth increase, the total energy of each spectrum appears to be equal. Thus, the spectra magnitudes appear to be decaying at a rate of $1/\sqrt{M}$ where $M$ is the modulation index. This decay suggests that spectral norm deviations should be relatively small, even if a signal consists of two tones whose frequencies are equivalent up to folding. As a result, such a sampling scheme should, in theory, possess a good RIP. To illustrate this, Fig. 7 shows spectrograms of a signal with exactly two tones (7.2 and 9.2 GHz) folded after both uniform sampling at 2 GHz and non-uniform sampling with a modulated linear chirp inducing a 100 MHz deviation. Because the two tones at 7.2 and 9.2 GHz fold to the same frequency, the two tones become indistinguishable after uniform sampling at 2 GHz. However, the linear chirp modulation induces each tone to take on a different slope in the bottom spectrogram, thus, making the two tones separable. The correlation of these two chirps is almost zero. As a result, we intuitively expect that a compressive sensing algorithm should be able to recover the tones from the non-uniform samples but not from the uniform samples.

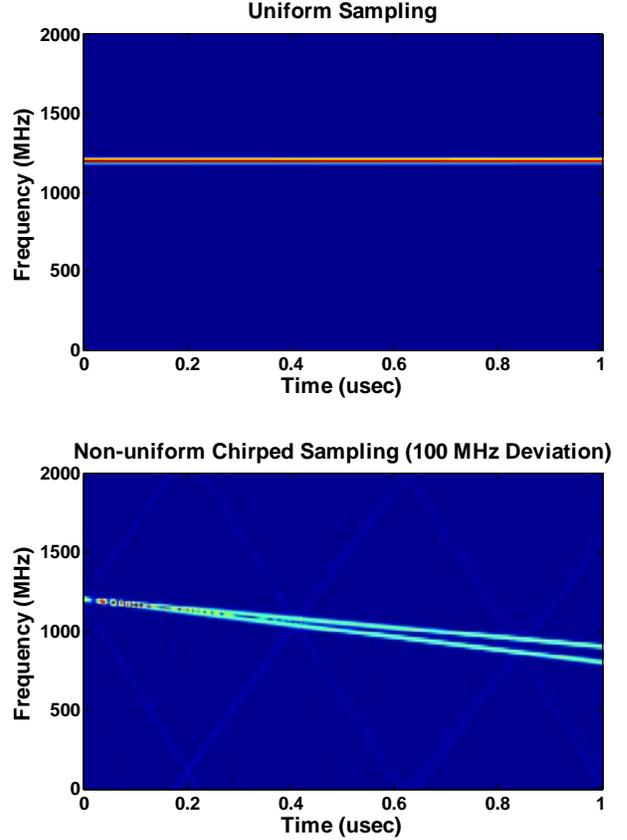

Fig. 7. Spectrograms of a folded two-tone signal after uniform and non-uniform chirped sampling.

To make this analysis more rigorous, we may proceed with a calculation similar to that in [7]. Assuming an extremely narrow pulse (i.e., a Dirac delta function), the non-uniform pulse train takes the form:

$$\tilde{p}(t) = \varphi'(t) \sum_k 2\pi \delta(\varphi(t) - 2\pi k)$$

where $\varphi(t) = \omega_{S1} t + \theta(t)$ is the expression from Equation (1). We use the identity $2\pi \sum_k \delta(t - 2\pi k) = \sum_k e^{jkt}$ and the fact that $\varphi'(t) = \omega_{S1} + \theta'(t)$ to rewrite the above as:

$$\tilde{p}(t) = (\omega_{S1} + \theta'(t)) \sum_k e^{jk(\omega_{S1} t + \theta(t))}$$

By realizing that $|\theta'(t)| \ll \omega_{S1}$, we may approximate the pulse train as:

$$\tilde{p}(t) = \omega_{S1} \sum_k e^{jk(\omega_{S1}t + \theta(t))}$$

When multiplied against a signal $x(t)$, the resulting frequency-domain output becomes:

$$Y(\omega) = X(\omega) * \omega_{S1} \sum_k \delta(\omega - k\omega_{S1}) * \mathcal{F}\{e^{jk\theta(t)}\}(\omega)$$
$$= X(\omega) * \omega_{S1} \sum_k \mathcal{F}\{e^{jk\theta(t)}\}(\omega - k\omega_{S1})$$
$$= \frac{\omega_{S1}}{2\pi} \int_{-\infty}^{\infty} \sum_k X(\xi) \mathcal{F}\{e^{jk\theta(t)}\}(\omega - \xi - k\omega_{S1}) \, d\xi$$

We define the $k$th spectrum of $x(t)$ to be

$$X_k(\omega) = \frac{1}{2\pi} \int_{-\infty}^{\infty} X(\xi) \mathcal{F}\{e^{jk\theta(t)}\}(\omega - \xi - k\omega_{S1}) d\xi$$

As stated earlier, we claim that each of these $X_k$s contains the same amount of energy:

**Theorem 3:** Given a square integrable signal $x(t)$, then

$$\|X_k(\omega)\|_2 = \|X(\omega)\|_2$$

for every spectrum $k$.

*Proof.* Observe that

$$\|X_k(\omega)\|_2 = \left\|\left(X * \mathcal{F}\{e^{jk\theta(t)}\}\right)(\omega - k\omega_{S1})\right\|_2$$
$$= \left\|\left(X * \mathcal{F}\{e^{jk\theta(t)}\}\right)(\omega)\right\|_2$$

where the second equality comes from the fact that $\mathcal{L}_2$ norms are shift-invariant. Now we use Parseval's relation to obtain:

$$\left\|\left(X * \mathcal{F}\{e^{jk\theta(t)}\}\right)(\omega)\right\|_2 = 2\pi \|x(t)e^{jk\theta(t)}\|_2$$
$$= 2\pi \|x(t)\|_2 = \|X(\omega)\|_2$$

as was to be showed.
□

This theorem, along with the help of Fig. 6, shows us that if the bandwidth of the $k$th spectrum of a narrow-band signal $x(t)$ increases to $kf_\text{dev}$ where $f_\text{dev}$ is the bandwidth of $e^{jk\theta(t)}$, then the average magnitude of this spectrum will be proportional to $\|X(\omega)\|_2/\sqrt{kf_\text{dev}}$ (depending on the normalization). Now in the case of a signal $x(t)$ consisting of exactly two tones, we can prove the following theorem whose proof is located in Appendix 1.

**Theorem 4:** Suppose we are given a discrete non-uniform sampling scheme induced by the rising voltage zero crossings of (1). Let $f_\text{res}$ be the atomic resolution of the discretized frequency space and let $f_\text{dev}$ be the bandwidth of $e^{jk\theta(t)}$. Further suppose there exists a constant $C$ such that

$$\left|\mathcal{F}\{e^{jk\theta(t)}\}(\omega)\right|^2 \leq \frac{C^2}{|k|f_\text{dev}} \int_{\mathfrak{B}_k} \left|\mathcal{F}\{e^{jk\theta(t)}\}(\xi)\right|^2 d\xi \quad (8)$$

for all $k$ and $\omega$ where $\mathfrak{B}_k$ is an interval representing the effective bandwidth of $e^{jk\theta(t)}$. We note that the integral in (8) is proportional to the root-mean-squared (RMS) magnitude of its integrand. In addition, suppose that $C\sqrt{f_\text{res}/f_\text{dev}} < 0.5$. Assuming the SNR of the system is sufficiently high, then the resulting discrete sampling scheme has an RIP constant of order two satisfying:

$$\delta_2 \leq C\sqrt{\frac{f_\text{res}}{f_\text{dev}}} = C\sqrt{\frac{1}{f_\text{dev}t_\text{atom}N}} = O\left(\frac{1}{\sqrt{N}}\right) \quad (9)$$

where $t_\text{atom}$ is the atomic step size of the discrete data and $N$ is the signal length.

An interesting observation is that as our discrete grid resolution gets finer, the above bound approaches:

$$\delta_2 \leq C\sqrt{\frac{1}{f_\text{dev}T_d}}$$

where $T_d$ is the signal duration. This is the extension to continuous signals we alluded to in Section 2. We can easily extend (9) to an upper bound on general RIP numbers by utilizing a result in [16] that states that

$$\delta_s \leq s\delta_2 \leq sC\sqrt{\frac{f_\text{res}}{f_\text{dev}}} = sC\sqrt{\frac{1}{f_\text{dev}T_d}} \quad (10)$$

As an illustrative example, using the discrete settings ($N = 1{,}000{,}000$, $K = 20{,}000$) consisting of a 200 MHz sample clock modulated with a linear chirp with period 100μs that induces a 10 MHz deviation, we can empirically calculate that $C = 1.21$ and $\sqrt{f_\text{res}/f_\text{dev}} = \sqrt{.01\text{MHz}/10\text{MHz}} = .0316$. Thus, we have:

$$\delta_2 \leq 1.21(.0316) = .0382$$

Recall that convex optimization will recover any 10μs signal consisting of exactly $s$ tones if $\delta_{2s} < \sqrt{2} - 1$. This will happen if

$$2s(.0316) < \sqrt{2} - 1$$

This implies that $s \leq 5$. In other words, assuming, a high SNR environment, convex optimization will recover any signal with five bin-centered tones regardless of dynamic range and how suboptimal their locations may be.

## VI. APPLICATION OF ORTHOGONAL MATCHING PURSUIT

In this section, we demonstrate how the popular compressive sensing algorithm Orthogonal Matching Pursuit (OMP) can effectively recover a superposition of several CW tones. Certainly, our theory dictates that OMP must be successful in a low noise environment.

According to Theorem 2, OMP will be able to recover any two tones if $\delta_3 < 1/(1 + \sqrt{2}) \approx 0.4142$. Assuming we sample a tone for 10μs at 200 MHz with linear chirp modulation ($f_{\text{dev}} = 10$ MHz), we find empirically that $C = 1.23$. Now according to Theorem 4 and Equation (10), it follows that

$$\delta_3 < 3.69\sqrt{\frac{1}{100}} = 0.369 < 0.4142$$

Thus, in low noise situations, OMP will recover *any* two tones from their samples regardless of any disparity in their strengths. Of course, RIP-based sufficient conditions tend to be fairly loose lower-bounds on performance: As we will demonstrate empirically, OMP can recover significantly more tones with high probability, even in the presence of noise and without the constraint that each tone falls into exactly one discrete frequency bin.

We performed an experiment where, for each sparsity level *s* from 3 to 60 in increments of 3, we generated 50 signals consisting of $s$ tones. The tones had uniform amplitude and random phase. In addition, the tones were *non-bin-centered* on our discrete frequency grid. Thus, strictly speaking, these signals were subject to spectral leakage and not sparse in frequency: they were only sparse with respect to the dictionary of all possible CW tones. We added various amounts of noise to these signals (20 dB, 10 dB, 0 dB SNR) and let OMP recover them. For any one particular signal, we declared a success if OMP recovered the frequencies of *all* the tones correctly. Otherwise, we had a failure. The results of this experiment are shown in Fig. 8.

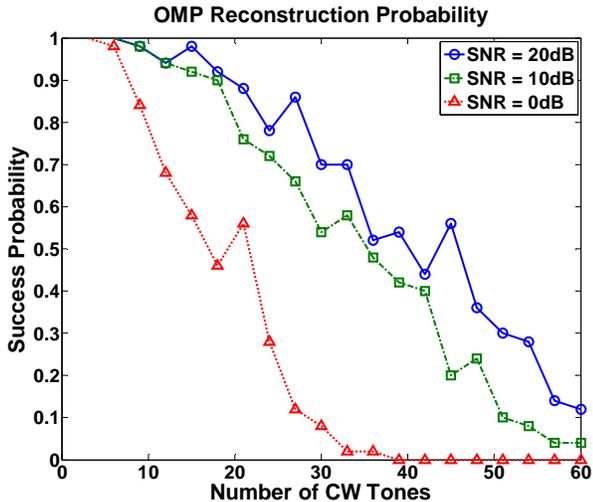

Fig. 8. Ability of OMP to recover a superposition of several CW tones in the presence of noise.

In the 10-20 dB range, up to 18 CW tones were recoverable assuming a 10% failure probability is acceptable. This is much better than the two recoverable tones guaranteed by RIP analysis.

An important question that needs to be addressed in order to make this discussion complete is how much noise can be present in order for a single tone to be identifiable. Assume we're given a signal consisting of a single tone corrupted with complex-valued white Gaussian noise, i.e.

$$x[n] = e^{j2\pi k_0 n/N} + w[n] \qquad (11)$$

where $w[n]$ is a white noise process with zero mean and variance $\sigma^2$. Then we can prove the following theorem. The proof is included in Appendix 2.

**Theorem 5:** Suppose we take *K* samples of the signal (11) according to the modulated sampling scheme. Then the probability that OMP will detect the tone at frequency $k_0$ from the given samples is bounded from below by:

$$p_D \geq \left[1 - \exp\left(\frac{-K(1-\delta_2)^2}{4\sigma^2}\right)\right]^N$$

where $\delta_2$ is the RIP constant in Equation (10).

As can be inferred from Fig. 7, an alternate method of determining the location of a tone from samples obtained from a linearly chirped sampling clock is to estimate the slope of the resulting chirp modulated onto the tone. Knowledge of this slope is equivalent to knowledge of the original tone's Nyquist zone. Assume we have access to *K* samples of the chirp

$$x_k = A\exp\left[j\left(2\pi\left(\frac{\alpha}{2}(\Delta k)^2 + f\Delta k\right) + \varphi\right)\right] + w_k \qquad (12)$$

where $w_k = a_k + jb_k$ and both $a_k$ and $b_k$ are independent and identically distributed Gaussian noise processes with zero mean and variance $\sigma^2/2$. The parameter $\Delta$ is the sampling step size. Our objective is to obtain an estimate $\hat{\alpha}$ of the chirp rate $\alpha$. We show in Appendix 3 that any estimator $\hat{\alpha}$ must satisfy the following Cramér-Rao bound:

$$\text{var}(\hat{\alpha}) \geq \frac{15\sigma^2}{A^2\pi^2\Delta^4 K(K+1)(2K+1)(3K^2+3K-1)}$$

Based on the CRB, we may calculate a lower bound on the probability of correct Nyquist zone selection. In Fig. 9, we compare this probability against that of OMP in Theorem 5 as well as empirically determined probabilities for CW tones with additive white noise (SNR = 10 dB) sampled at approximately 200 MHz. To empirically estimate the NZ estimation probabilities, for any given number of samples, we generated 50 signals each consisting of a single tone at a randomly chosen frequency plus additive noise. The experimental probability of NZ estimation is the proportion of times that OMP detected a frequency corresponding to the correct frequency (or Nyquist zone). For both the CRB and the empirical probability calculations, it is assumed that any given tone lies within the first 20 Nyquist zones. The RF sampling clock used for this experiment linearly transitions from 200 MHz to 210 MHz over a 10 μs period of time, i.e. a chirp slope of 1 MHz per μs.

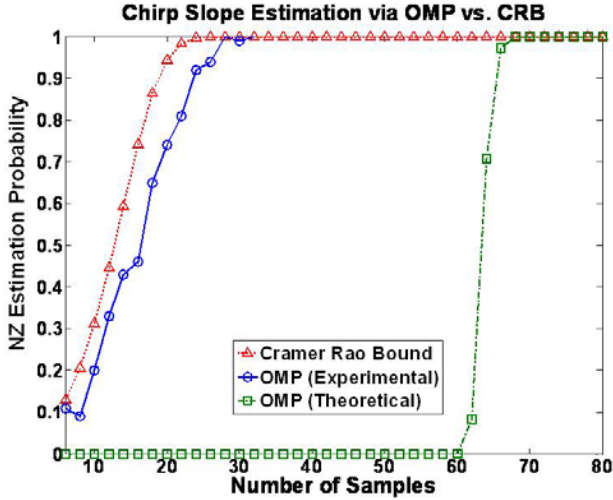

Fig. 9. Probability of Nyquist zone estimation (Cramer Rao Bound & OMP)

Now it may appear inappropriate to compare a Nyquist zone identification probability with a tone identification probability; however, the comparison is valid because if OMP correctly identifies a tone, then the Nyquist zone is also correctly identified. On the other hand, if a tone is incorrectly selected, then the probability of selecting the correct Nyquist zone is also very low (approximately the reciprocal of the total number of Nyquist zones).

The OMP experiment probability curve closely hugs the curve derived from the CRB. On the other hand, the OMP lower bound curve lags the other two curves by a great margin. As a result, we see that OMP is a statistically powerful tool for estimating the Nyquist zone of a signal.

## VII. EMPIRICAL EVALUATION OF SPECTRAL NORM DEVIATION

The restricted isometry property is an extremely useful tool in the analysis of compressive sensing methods. Unfortunately, the task of estimating RIP constants is a very arduous endeavor that typically does not culminate in extremely significant results. The STRIP does offer some improvement; however, the recoverable sparsity estimates are still not great. Oftentimes, compressive sensing algorithms are able to recover many signals that would otherwise not be guaranteed to be recoverable by any sort of restricted isometry condition. As a result, it is useful to venture beyond the current limitations of rigorous mathematical analysis and explore the spectral norm deviations of signals that arise in practice.

In this section, we empirically explore the effective RIP of a modulated sampling scheme. Let $x$ be a wide-band signal consisting of $s$ bin-centered tones. We discretize $x$ using an atomic time-resolution of 10 ps. This corresponds to a 100 GHz atomic sampling rate. We will sample with a clock running at 2.0 GHz that carries a modulated sinusoid with period 5μs and various amplitudes (frequency deviations) including 0 MHz (uniform sampling), 10 MHz, and 100 MHz. We performed the following test for each case: For sparsity level from $s = 400$ to $s = 4000$ in increments of 400, we generated 100 experimental signals with exactly $s$ bin-centered tones. The amplitudes of these tones are independent Gaussian random numbers with zero mean and unit variance. For each sample of 100 signals, we calculated the maximum spectral norm deviation (5) using each modulation bandwidth and claim that this is a measure of the effective RIP number of the respective sampling schemes. In other words, it is unlikely that one will generate signals that yield spectral norm deviations larger than those shown in Fig. 10.

Interestingly enough, the effective RIP numbers are all linearly related to the sparsity level but seem to have no relationship to the degree of non-uniformity introduced. In other words, if we neglect the unlikely event of a folding collision, all Fourier sampling schemes behave similarly from a compressive sensing perspective. This is consistent with the prediction of the statistical RIP discussed in Section IV.

Of course, this analysis assumes that we are working with extremely narrow-band pure bin-centered tones in a noiseless environment. Thus, the estimates of Fig. 10 are highly optimistic and should not be expected in reality. However, as we have shown theoretically and empirically, compressive sensing methods are highly effective at recovering several tones with high probability even in the presence of reasonable noise.

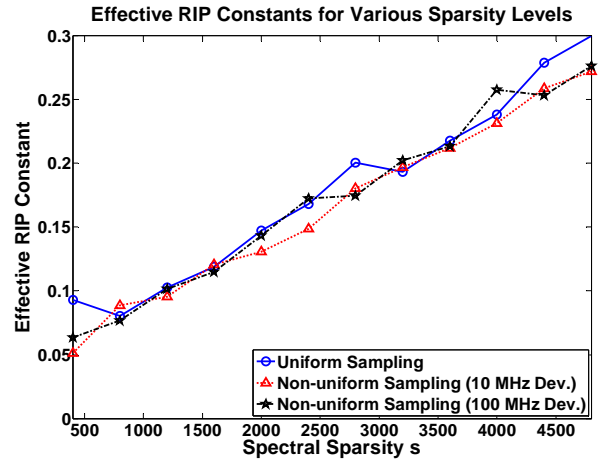

Fig. 10. Empirical evaluation of spectral norm deviations.

## VIII. CONCLUSION

The primary contribution of this work is to analyze modulated sampling schemes such as the Nyquist Folding Receiver within a compressive sensing framework. Because such sampling schemes spread out the aliased spectra of a signal, it follows from an energy conservation argument that it must attenuate them spectrally as well. Thus, while structured enough to preserve the signal spectral content, modulated sampling possesses enough non-uniformity to also enjoy strong restricted isometry properties. As a result, compressive sensing algorithms can be highly effective in terms of recovering signals sampled using this methodology. We have estimated the RIP constants for modulated sampling schemes both empirically and theoretically. Moreover, we have validated the use of Orthogonal Matching Pursuit (OMP) as an effective recovery algorithm, even in the presence of significant noise. Thus, the modulated sampling scheme, together with CS reconstruction techniques, form a powerful approach with respect to the blind estimation of wide-band signals with sparsely located narrow-band components.

ACKNOWLEDGEMENTS

We would like to thank Paul Deignan and Jarvis Haupt for their discussions and sharing of previous work that enabled us to perform this analysis. We thank Rob Nowak for his helpful comments regarding the content of this paper. A portion of this work was sponsored by the Air Force Research Laboratory through the DARPA Analog-to-Information Program.

APPENDIX 1: PROOF OF THEOREM 4

Assuming very little noise in the system, the worst norm deviation will occur for a signal $x$ consisting of tones at frequencies $f_1$ and $f_2$ where $f_1$ and $f_2$ are equivalent to folding and sit two Nyquist zones apart, i.e., their modulation indices $M_1$ and $M_2$ differ by one. Let the amplitude of these tones be $A_1$ and $A_2$ respectively. Equation (8) and Theorem 3 together imply that $\Phi^*\Phi\hat{x}$ will have a value in the Euclidean ball $\mathbb{B}_{|\alpha A_2|}(A_1)$ at position $f_1$ where $\alpha = C\sqrt{f_{\text{res}}/f_{\text{dev}}}$. Call this point $A_1 + \alpha_1 A_2$. Similarly, at position $f_2$, $\Phi^*\Phi\hat{x}$ will have a value in the Euclidean ball $\mathbb{B}_{|\alpha A_1|}(A_2)$ which we will denote $A_2 + \alpha_2 A_1$. Now the term in the square root of the spectral norm deviation (5) will become:

$$\frac{|A_1 + \alpha_1 A_2|^2 + |A_2 + \alpha_2 A_1|^2}{|A_1|^2 + |A_2|^2} \quad (13)$$

$$= \frac{|A_1|^2 + |A_2|^2 + |\alpha_2|^2|A_1|^2 + |\alpha_1|^2|A_2|^2 + 2\text{Re}[\alpha_1 A_1 A_2] + 2\text{Re}[\alpha_2 A_1 A_2]}{|A_1|^2 + |A_2|^2}$$

$$\leq 1 + \alpha^2 + \frac{4\alpha|A_1||A_2|}{|A_1|^2 + |A_2|^2}$$

where the last line utilizes the Cauchy-Schwartz inequality. One can use elementary calculus to show that this quantity is maximized when $|A_1| = |A_2|$. This gives us that the spectral norm deviation must always be bounded from above by

$$\delta_2(x) \leq \sqrt{1 + \alpha^2 + 2\alpha} - 1 = \alpha$$

We next calculate a lower bound for (13). Assume without loss of generality that $|\alpha_1| \geq |\alpha_2|$.

$$\frac{|A_1|^2 + |A_2|^2 + |\alpha_2|^2|A_1|^2 + |\alpha_1|^2|A_2|^2 + 2\text{Re}[\alpha_1 A_1 A_2] + 2\text{Re}[\alpha_2 A_1 A_2]}{|A_1|^2 + |A_2|^2}$$

$$\geq \frac{|A_1|^2 + |A_2|^2 + |\alpha_2|^2|A_1|^2 + |\alpha_2|^2|A_2|^2 + 2\text{Re}[\alpha_1 A_1 A_2] + 2\text{Re}[\alpha_2 A_1 A_2]}{|A_1|^2 + |A_2|^2}$$

$$\geq 1 + |\alpha_2|^2 + \frac{2\text{Re}[\alpha_1 A_1 A_2] + 2\text{Re}[\alpha_2 A_1 A_2]}{|A_1|^2 + |A_2|^2}$$

We apply the Cauchy-Schwartz inequality to further bound this by:

$$\geq 1 + |\alpha_2|^2 - \frac{2|\alpha_1||A_1||A_2| + 2|\alpha_2||A_1||A_2|}{|A_1|^2 + |A_2|^2}$$

This quantity is minimized when $|A_1| = |A_2|$. Making this substitution yields:

$$1 + |\alpha_2|^2 - |\alpha_1| - |\alpha_2|$$

Now use the fact that $|\alpha_2| \leq |\alpha_1| \leq \alpha \leq 0.5$ and the fact that the function $f(\alpha) = \alpha^2$ has slope less than 1 for $\alpha \leq 0.5$ to deduce that:

$$1 + |\alpha_2|^2 - |\alpha_1| - |\alpha_2| \geq 1 + |\alpha_1|^2 - |\alpha_1| - |\alpha_1|$$

which implies that the spectral norm deviation still satisfies:

$$\delta_2(x) \leq \left|\sqrt{1 + \alpha_1^2 + 2\alpha_1} - 1\right| = |-\alpha_1| \leq \alpha$$

The main statement of the theorem now immediately follows.

APPENDIX 2: PROOF OF OMP PROBABILITY OF DETECTION

Let $x_s[n]$ be the sampled version of $x$ where $x_s[n] = x[n]$ at sample points and zero otherwise. Let $X_s[k]$ be its DFT. Similarly define the sampled noise signal $w_s[n]$ and its DFT $W_s[k]$. We can easily check that $\mathbb{E}[W_s[k]] = 0$ and $\text{var}[W_s[k]] = \sigma^2 K$. We can also check that $X_s[k_0] = K + W_s[k_0]$. Every other value of $X_s[k]$ will be the sum of a quantity that is bounded by $\delta_2$ in magnitude and a noise term $W_s[k]$. Now since $W_s[k]$ is complex valued with independent components, it follows that $\text{Re}(W_s[k]) \sim \mathcal{N}(0, \sigma\sqrt{K/2})$ and $\text{Im}(W_s[k]) \sim \mathcal{N}(0, \sigma\sqrt{K/2})$, and $|W_s[k]|$ is Rayleigh distributed. OMP will recover $k_0$ if we have that $|W_s[k]| < K(1 - \delta_2)/2$ for every $k$. Thus, we observe that the probability of detection satisfies

$$p_D \geq \mathbb{P}\left(\bigcap_{k=1}^{N}\{|W_s[k]| < K(1 - \delta_2)/2\}\right)$$

$$= \mathbb{P}(|W_s[1]| < K(1 - \delta_2)/2)^N$$

$$= \left(1 - \exp\left(\frac{-K(1 - \delta_2)^2}{4\sigma^2}\right)\right)^N$$

where the last line is obtained by utilizing the cumulative distribution function for the Rayleigh distribution.

APPENDIX 3: DERIVATION OF CRAMÉR-RAO BOUND FOR CHIRP ESTIMATION

We can separate (12) into real and imaginary components which gives us:

$$x_k = A\cos\left[\left(2\pi\left(\frac{\alpha}{2}(\Delta k)^2 + f\Delta k\right) + \varphi\right)\right] + a_k$$
$$+ jA\sin\left[\left(2\pi\left(\frac{\alpha}{2}(\Delta k)^2 + f\Delta k\right) + \varphi\right)\right] + b_k$$

We next express the joint distribution of the samples $x_k$ as:

$$p(x_k|\alpha) = \frac{1}{(\pi\sigma^2)^N}\exp\left(\frac{1}{\sigma^2}\sum_{i=1}^{K}(x_i^R - A\cos[\pi\alpha(\Delta i)^2 + 2\pi f\Delta i + \varphi])^2 \right.$$
$$\left. + \frac{1}{\sigma^2}\sum_{i=1}^{K}(x_i^I - A\sin[\pi\alpha(\Delta i)^2 + 2\pi f\Delta i + \varphi])^2\right)$$

where the $x_i^R$s and the $x_i^I$s are the respective real and imaginary parts of the samples $x_i$. We take the logarithm of this density function to obtain the likelihood function:

$$L(x_k|\alpha) = \frac{1}{\sigma^2}\sum_{i=1}^{K}(x_i^R - A\cos[\pi\alpha(\Delta i)^2 + 2\pi f\Delta i + \varphi])^2$$
$$+ \frac{1}{\sigma^2}\sum_{i=1}^{K}(x_i^I - A\sin[\pi\alpha(\Delta i)^2 + 2\pi f\Delta i + \varphi])^2 - N\ln(2\pi\sigma^2)$$

The second derivative of this quantity with respect to the parameter $\alpha$ can easily be shown to be:

$$\frac{\partial^2 L}{\partial \alpha^2} = \frac{2}{\sigma^2}\sum_{i=1}^{K}x_i^R A\,\cos[\pi\alpha(\Delta i)^2 + 2\pi f\Delta i + \varphi]\pi^2(\Delta i)^4$$
$$+ \frac{2}{\sigma^2}\sum_{i=1}^{K}x_i^I A\,\sin[\pi\alpha(\Delta i)^2 + 2\pi f\Delta i + \varphi]\pi^2(\Delta i)^4$$

The Fisher information is then calculated as:

$$I(\alpha) = \mathbb{E}\left(\frac{\partial^2 L}{\partial \alpha^2}\right) = \frac{2}{\sigma^2}\sum_{i=1}^{K}A^2\cos^2[\pi\alpha(\Delta i)^2 + 2\pi f\Delta i + \varphi]\pi^2(\Delta i)^4$$
$$+ \frac{2}{\sigma^2}\sum_{i=1}^{K}A^2\sin^2[\pi\alpha(\Delta i)^2 + 2\pi f\Delta i + \varphi]\pi^2(\Delta i)^4$$

Finally, we use the trigonometric identity $\cos^2\theta + \sin^2\theta = 1$ to obtain:

$$I(\alpha) = \frac{1}{\sigma^2}\sum_{i=1}^{K}A^2\pi^2(\Delta i)^4$$
$$= \frac{A^2\pi^2\Delta^4 K(K+1)(2K+1)(3K^2+3K-1)}{15\sigma^2}$$

Thus, any estimator $\hat{\alpha}$ of $\alpha$ must satisfy the following Cramér-Rao bound:

$$\mathrm{var}(\hat{\alpha}) \geq \frac{15\sigma^2}{A^2\pi^2\Delta^4 K(K+1)(2K+1)(3K^2+3K-1)}$$

as was to be shown. It should be noted that this result is slightly different than that of [24] which assumes that the chirp sampling scheme is centered.